\pdfoutput=1

\documentclass[sigconf,natbib=true,anonymous=false]{acmart}

\setcopyright{none}
\renewcommand\footnotetextcopyrightpermission[1]{}

\usepackage[T1]{fontenc}

\usepackage[hang,flushmargin]{footmisc}

\usepackage{url}            
\usepackage{booktabs}       
\usepackage{amsfonts}       
\usepackage{nicefrac}       
\usepackage{microtype}      
\usepackage{amsmath}
\usepackage{enumitem}
\usepackage{graphicx}       
\usepackage{caption}
\usepackage{subcaption}

\title{Anserini Gets Dense Retrieval:\\ Integration of Lucene's HNSW Indexes}

\author{Xueguang Ma$^{1}$, Tommaso Teofili$^{2}$, Jimmy Lin$^1$}

\affiliation{\vspace{0.1cm}$^1$ David R. Cheriton School of Computer Science, 
University of Waterloo \country{Canada}\\
$^2$ Roma Tre University \country{Italy}}

\begin{document}

\renewcommand{\shortauthors}{}
\pagestyle{empty}

\begin{abstract}
Anserini is a Lucene-based toolkit for reproducible information retrieval research in Java that has been gaining traction in the community.
It provides retrieval capabilities for both ``traditional'' bag-of-words retrieval models such as BM25 as well as retrieval using learned sparse representations such as SPLADE.
With Pyserini, which provides a Python interface to Anserini, users gain access to {\it both} sparse and dense retrieval models, as Pyserini implements bindings to the Faiss vector search library alongside Lucene inverted indexes in a uniform, consistent interface.
Nevertheless, hybrid fusion techniques that integrate sparse and dense retrieval models need to stitch together results from two completely different ``software stacks'', which creates unnecessary complexities and inefficiencies.
However, the introduction of HNSW indexes for dense vector search in Lucene promises the integration of both dense and sparse retrieval within a single software framework.
We explore exactly this integration in the context of Anserini.
Experiments on the MS MARCO passage and BEIR datasets show that our Anserini HNSW integration supports (reasonably) effective and (reasonably) efficient approximate nearest neighbor search for dense retrieval models, using only Lucene.
\end{abstract}
\maketitle

\section{Introduction}

One important recent development in neural information retrieval is that dense retrieval models based on the so-called bi-encoder (or dual-encoder) architecture require an entirely different ``software stack'' for efficient top-$k$ retrieval~\cite{dpr, faiss}.
These models take advantage of deep neural models (typically, pretrained transformers~\cite{attention}) to encode both queries and documents into dense representation vectors~\cite{dpr, ance, tct_colbert, cocondenser}, where ranking is framed as a nearest neighbor search problem in vector space.
Today, efficient execution in an operational setting is usually performed using software libraries such as Faiss~\cite{faiss}, which implements hierarchical navigable small-world network (HNSW) indexes~\cite{HNSW}.
This query evaluation infrastructure is quite distinct and largely incompatible with sparse retrieval (e.g., BM25) using inverted indexes, which has been the mainstay of search systems dating back many decades~\cite{robertson2009bm25}.

Despite the effectiveness of dense retrieval models, we do not anticipate inverted indexes becoming obsolete in the foreseeable future.
Thus, achieving state-of-the-art results today requires maintaining two separate ``software stacks'':\ one for dense retrieval and another for sparse retrieval.
For dense retrieval, the Faiss library~\cite{faiss} has emerged as the most popular solution, and for sparse retrieval, the Lucene search library (which underlies popular search platforms such as Elasticsearch and Solr) remains the go-to solution.
While there exist integrated systems that manage both sparse and dense retrieval models such as Vespa,
custom implementations of hybrid fusion combining Faiss and Lucene results is a common architectural pattern~\cite{Ma_etal_arXiv2021_DPR, tct_colbert}.
In real-world applications, such a design creates non-trivial engineering challenges.
For example, indexes need to be kept up-to-date as documents are added and removed from the collection; retrieval requires scheduling two different types of queries.
While not insurmountable, these issues add complexity to deployed systems and friction during system development.

A new feature in the latest major release of Lucene (version 9) is the implementation of HNSW indexes~\cite{HNSW}.
This means that Lucene now provides support for {\it both} inverted and HNSW indexes.
However, these indexes remain under-unexplored using formal text retrieval benchmarks such as MS MARCO passage~\cite{msmarco}.
How do effectiveness and efficiency of Lucene's HNSW implementation compare to that of Faiss?
This question remains unanswered.

To aid in this exploration, we integrate Lucene's HNSW indexes into the open-source Anserini IR toolkit~\cite{anserini} alongside existing inverted indexes.
With this addition, Anserini provides a unified interface for dense and sparse retrieval.
We compare this ``experience'' with the previous approach that requires Pyserini, a Python interface to Anserini that splits dense and sparse retrieval across Faiss and Lucene, respectively~\cite{lin2021pyserini}.
We provide empirical evaluations in terms of effectiveness and various aspects of efficiency---query throughput, indexing throughput, and index size---on modern retrieval test collections, MS MARCO passage~\cite{msmarco} and BEIR~\cite{beir}.

\smallskip
\noindent {\bf Contributions.} We view this work as having three contributions:

\begin{itemize}[leftmargin=*]

\item We provide integration of Lucene's HNSW indexing and search capabilities in Anserini, which now offers a ``one-stop shop'' for reproducible research on common IR test collections for dense as well as sparse representations.

\item Due to an implementation choice in Lucene, proper use of HNSW indexes requires training new models that use cosine similarity as the similarity metric (instead of the more common inner product).
We have trained such a dense retrieval model, which we share on the Huggingface Model Hub.

\item With new capabilities in Anserini and a new dense retrieval model, we provide a fair evaluation of tradeoffs between result quality, time (indexing and query evaluation throughput), and space (index size), comparing Lucene vs.\ Faiss.

\end{itemize}

\noindent At a high level, this work brings two practical benefits.
For the researcher:\ integration of Lucene HNSW indexes into Anserini simplifies the ``research toolchain'' and streamlines experimental development.
For the practitioner:\ experimental results in this paper provide a fair comparison between the HNSW implementations in Lucene and Faiss, which provides concrete guidance on the time/space/quality tradeoffs that are important in deployed production systems.

\section{Background and Related Work}

Inverted indexes have been the ``go-to'' solution for the top-$k$ retrieval problem based on ``bag-of-words'' retrieval models such as BM25 for many decades~\cite{robertson2009bm25, anserini, drqa}.
However, one of the limitations of traditional inverted indexes is that they are unable to capture semantic relationships.
Recently, dense retrieval models based on vector representations generated by transformers~\cite{attention}
have emerged as a popular alternative~\cite{dpr, tct_colbert}.
In contrast to document vectors that are generated by traditional bag-of-words models, dense representations aim to capture semantic information contained in texts.

While dense retrieval models have proven to be effective, they are not as efficient as traditional inverted indexes when it comes to fast exact match algorithms~\cite{coil, dpr-compress}.
Approximate nearest neighbor methods offer a way to speed up dense retrieval models~\cite{faiss} while retaining their effectiveness.
One of the most popular algorithms for efficient nearest neighbor search is the hierarchical navigable small-world (HNSW) graph~\cite{HNSW}. 
An HNSW index comprises a multi-layered graph where each layer holds a proximity graph for a subset of the stored vectors.
During the search process, the algorithm starts from the top layer and descends through the graph layers to locate the approximate $k$-nearest neighbors;
HNSW indexes can accomplish this with a high degree of accuracy, while also being relatively efficient and scalable.

Despite the popularity of dense retrieval models, lexical retrieval is still necessary in practice, we argue.
This claim is based on at least two observations:\ (1) First, dense retrieval models still exhibit issues such as out-of-domain generalization~\cite{beir}.
(2) Sparse learned representations demonstrate competitive effectiveness compared to dense retrieval models~\cite{splade}.
Generally, results from the literature consistently report that hybrid retrieval architectures---combining dense and sparse results using fusion techniques such as linear score weighting or reciprocal rank fusion---are more effective than their individual constituents~\cite{unicoil}.

Despite the existence of several dense retrieval libraries such as Faiss~\cite{faiss}, there is still a need for a comprehensive {\it single} framework that can facilitate both dense and sparse retrieval.
Recently, Lucene accomplished this, and we have added bindings to Lucene's HNSW implementation in Anserini.
This integration bridges the software deployment gap between dense and sparse retrieval models, providing a (potentially) unified solution for researchers and practitioners, perhaps enabling more extensive collaborations between academia and industry~\cite{Devins_etal_WSDM2022}. 

\section{Implementation}

In this section, we describe the integration of HNSW indexing and search in Anserini.
Following the same design pattern (exposed via \texttt{IndexCollection} and \texttt{SearchCollection} classes) for lexical retrieval based on inverted indexes, we introduce the class
\texttt{IndexHnswDenseVectors} for indexing and the corresponding class \texttt{SearchHnswDenseVectors} for search.

For indexing, we implemented the \texttt{IndexHnswDenseVectors} class, which takes a collection of dense vectors (in \texttt{jsonl} format) and creates an HNSW index using Lucene.
The index is constructed in a segmented fashion, where each segment represents a subset of the input collection; orchestration of the segments is handled by Lucene behind the scenes. 
Same as the description in the original HNSW paper~\cite{HNSW}, our indexing implementation exposes (i) the \texttt{efConstruction} parameter for selecting the number of nearest neighbor candidates to track while searching the graph for each newly inserted node and (ii) the \texttt{M} parameter to control the number of bi-directional links created for a new node during index construction.
The \texttt{IndexHnswDenseVectors} class provides options for multi-threaded indexing, where the number of threads is specified by the user.
Additionally, Anserini exposes an \texttt{optimize} option to control whether all index segments should be merged into a single segment once all documents have been ingested.
Changing the parameters has the following effects:

\begin{itemize}[leftmargin=*]

    \item \texttt{efConstruction}: Increasing this value enhances retrieval effectiveness at the cost of longer indexing times.
    \item \texttt{M}: Increasing this value enhances retrieval effectiveness at the cost of longer indexing times and bigger indexes.
    \item \texttt{threads}: Increasing this value speeds up indexing but consumes more resources.
    \item \texttt{optimize}: This option will yield  faster search at the cost of additional time to merge the index segments.
\end{itemize}

\noindent Overall, our implementation provides flexibility to explore the tradeoffs of different settings.
The command below is an example of using Anserini to build an HNSW index:

\smallskip
\begin{small}
\begin{verbatim}
target/appassembler/bin/IndexHnswDenseVectors \
  -collection JsonDenseVectorCollection \
  -input collections/msmarco-passage-cos-dpr-distil \
  -index indexes/hnsw.msmarco-passage-cos-dpr-distil/ \
  -generator LuceneDenseVectorDocumentGenerator \
  -threads 16 \
  -M 16 -efC 100
\end{verbatim}
\end{small}

\noindent To search dense vectors, we provide the \texttt{SearchHnswDenseVectors} class, which takes query vectors and searches the HNSW index for the top-$k$ most similar vectors. 
The \texttt{efSearch} parameter is exposed to control the number of candidate neighbors that are explored during search time.
Increasing \texttt{efSearch} increases effectiveness but increases query latency as well.
Similar to indexing, there are effectiveness--efficiency tradeoffs for search.
The search class also provides options for multi-threaded searching; the current implementation provides inter-query parallelism, where multiple queries are processed in parallel.

Overall, with this class, users have the ability to perform search under different performance tradeoff scenarios.
The command below is an example of using Anserini to search over the HNSW index created above:
\smallskip
\begin{small}
\begin{verbatim}
target/appassembler/bin/SearchHnswDenseVectors \
  -index indexes/hnsw.msmarco-passage-cos-dpr-distil \
  -topics topics.msmarco-passage.cos-dpr-distil.jsonl.gz \
  -topicreader JsonIntVector \
  -querygenerator VectorQueryGenerator \
  -output runs/run.cos-dpr-msmarco-dev.trec \
  -topicfield vector \
  -threads 16 \
  -hits 1000 \
  -efSearch 1000
\end{verbatim}
\end{small}

\noindent Once retrieval finishes, the results are stored in a text file that can be evaluated using standard tools such as \texttt{trec\_eval}.

\section{A New Dense Retrieval Model}

Dense retrieval models for the MS MARCO passage dataset, such as ANCE~\cite{ance}, TAS-B~\cite{tasb}, and TCT\_ColBERT~\cite{tct_colbert}, are typically trained using dot product similarity.
However, the Lucene HNSW implementation requires that the dense representations have an L2 norm of one, which means that we need a model that is trained to generate vector representations for cosine similarity.
In this work, we use ``best practices'' such as distillation from a cross encoder~\cite{rocketqa} and hard negative mining~\cite{ance, rocketqa, cocondenser} to train a dense retrieval model that is effective for cosine similarity and adaptable to our Anserini HNSW implementation.
We call this model cosDPR-distil.

For our model, we first train a dense retriever using the MS MARCO passage training data with BM25 hard negatives.
We use the model to mine hard negative passages for queries in the training set and combine the hard negative passages with the original BM25 hard negatives as augmented training data.
Then, we use the augmented training data to train a cross-encoder reranker.
We finally use the reranker to teach a dense retriever student, which is the final cosDPR-distil model, with knowledge distillation over the augmented training data.

We used Tevatron~\cite{tevatron} to train our model following the process outlined above on a server equipped with 8 $\times$ A6000 GPUs.
Specifically, we initialized the cross-encoder teacher with the Electra-large model~\cite{clark2020electra} and trained it using a batch size of 16 $\times$ 16 (each batch contains 16 queries and each query contains 1 positive passage and 15 negative passages)\ for 6 epochs.
For the dense retriever student, we initialized the model with Co-Condenser~\cite{cocondenser} and trained it using a batch size of 16 $\times$ 16 for 10 epochs.
During the process of hard negative mining, we randomly selected 30 documents for each query from the top 200 retrieval results of the dense retriever that were not labeled as relevant.

\section{Experiment Setup}

We evaluated our Anserini HNSW integration on the MS MARCO passage and BEIR datasets.

\begin{itemize}[leftmargin=*]

    \item MS MARCO passage is a web search dataset that consists of a training set with approximately 500k relevant query--passage pairs and a corpus with 8.8 million passages. For evaluation, we used the development set, which includes 6,980 queries. Our main objective is to compare the efficiency and effectiveness of the HNSW implementation in Lucene and Faiss. Effectiveness is measured by two metrics, MRR@10 and Recall@1k, while efficiency is measured in terms of QPS (queries per second).

    \item BEIR is a collection of datasets specifically created for zero-shot text retrieval evaluation. This collection consists of 18 datasets covering a wide range of domains. We used BEIR to analyze out-of-domain effectiveness generalization. We ran our experiments on the 13 publicly accessible datasets.

\end{itemize}

\noindent To evaluate the time required for index construction and search efficiency for both Lucene and Faiss HNSW indexes, we used a Google Cloud VM with the \texttt{e2-standard-16} configuration, which contains 16 vCPU cores, 64 GB RAM and 1TB SSD.
Experiments were performed with Lucene 9.4.2.
All experiments with Faiss were performed in Pyserini.
In all cases, we used pre-encoded queries and documents---that is, our timings do {\it not} include neural inference.

\section{Results}

\begin{table}[t]
\centering
\begin{tabular}{l|cc}
\toprule
& MRR@10 & Recall@1k\\
\midrule
BM25 & 0.184 & 0.853 \\
ANCE & 0.330 & 0.959 \\
TAS-B & 0.344 & 0.977 \\
CoCondenser & 0.382 & 0.984 \\
\midrule
cosDPR-distil (Flat) & 0.390 & 0.980  \\
cosDPR-distil (HNSW) & 0.389 & 0.974 \\
\bottomrule
\end{tabular}
\vspace{0.25cm}
\caption{Comparing our cosDPR-distil model with existing retrieval models on the MS MARCO passage ranking dev set.}
\label{table:marco}
\vspace{-0.75cm}
\end{table}

\begin{table*}[t]
\centering
\begin{tabular}{l|cccccccc}
\toprule
& & & & & \multicolumn{2}{c}{cosDPR-distil} \\
& BM25 & DPR & ANCE & TASB & Flat & HNSW \\
\midrule
TREC-COVID        & 0.656 & 0.332 & 0.654 & 0.481 & 0.497 & 0.497 \\
NFCorpus          & 0.325 & 0.189 & 0.237 & 0.319 & 0.319 & 0.319 \\
NQ                & 0.329 & 0.474 & 0.446 & 0.463 & 0.468 & 0.467 \\
HotpotQA          & 0.603 & 0.391 & 0.456 & 0.584 & 0.517 & 0.510 \\
FiQA-2018         & 0.236 & 0.112 & 0.295 & 0.300 & 0.238 & 0.237 \\
ArguAna           & 0.414 & 0.175 & 0.415 & 0.429 & 0.472 & 0.472 \\
Tóuche-2020 (v2)  & 0.367 & 0.127 & 0.240 & 0.162 & 0.180 & 0.177 \\
Quora             & 0.789 & 0.248 & 0.852 & 0.835 & 0.859 & 0.858 \\
DBPedia           & 0.313 & 0.263 & 0.281 & 0.384 & 0.375 & 0.371 \\
SCIDOCS           & 0.158 & 0.077 & 0.122 & 0.149 & 0.143 & 0.143 \\
FEVER             & 0.753 & 0.562 & 0.669 & 0.700 & 0.591 & 0.580 \\
Climate-FEVER     & 0.213 & 0.148 & 0.198 & 0.228 & 0.196 & 0.197 \\
SciFact           & 0.665 & 0.318 & 0.507 & 0.643 & 0.561 & 0.561\\
\midrule
Avg.nDCG@10 & 0.448 & 0.263 & 0.413 & 0.416 & 0.417 & 0.415 \\
\bottomrule
\end{tabular}
\vspace{0.25cm}
\caption{Comparing our cosDPR-distil model with existing retrieval models on BEIR.}
\label{table:beir}
\vspace{-0.25cm}
\end{table*}

\begin{table*}[t]
\centering
\begin{tabular}{ll|rrrrrllr}
\toprule
 & Index & efC & Threads & Optimize & Time (h) & Size (GB) & MRR@10 & Recall@1k & QPS \\
\midrule
(1) & Anserini & 100       & 1       &  no   &  13.6  & 27  & 0.3887 & 0.9737 & 3.6 \\
(2) & Anserini & 100       & 16      &  no   &  2.9   & 27  & 0.3890 & 0.9743 & 1.5 \\
(3) & Anserini & 1000      & 16      &  no   &  19.6  & 27  & 0.3894 & 0.9784 & 1.5  \\
(4) & Anserini & 100       & 1       &  yes  &  21.6  & 27  & 0.3842 & 0.9600 & 35.8 \\
(5) & Anserini & 100       & 16      &  yes  &  10.7  & 27  & 0.3795 & 0.9546 & 31.1 \\
(6) & Anserini & 1000      & 16      &  yes  &  96.5  & 27  & 0.3869 & 0.9737 & 23.6 \\
\midrule
(7) & Faiss FlatIP & -       & -       &  -    &  -    & 26  & 0.3896 & 0.9796 & 0.4 \\
(8) & Faiss HNSW   & 100     & 1       &  -    &  3.8  & 27  & 0.3806 & 0.9522 & 60.5\\
(9) & Faiss HNSW   & 100     & 16      &  -    &  0.3  & 27  & 0.3809 & 0.9584 & 60.0 \\
(10) & Faiss HNSW  & 1000    & 16      &  -    &  3.3  & 27  & 0.3866 & 0.9730 & 44.3 \\
\bottomrule
\end{tabular}
\vspace{0.25cm}
\caption{HNSW search with different indexing configurations. We set \texttt{M}=16 for indexing and \texttt{efSearch}=1000 for search.}
\label{table:search-with-different-indexing-configs}
\vspace{-0.5cm}
\end{table*}

\paragraph{Evaluation of model effectiveness}
In Tables \ref{table:marco} and \ref{table:beir}, we show the effectiveness of our cosDPR-distil model compared to existing representative dense retrieval models on the MS MARCO passage and BEIR datasets.
For our model, we provide results using both ``flat'' indexes (i.e., brute-force comparisons) and HNSW indexes.
The HNSW index here was built with \texttt{M}=16, \texttt{efC}=100, using a single thread with optimization.
Results for the other models are with ``flat'' indexes.

For the MS MARCO passage dataset (in-domain), our model demonstrates competitive effectiveness compared to other dense retrieval baselines.
Under the zero-shot setting on BEIR (out-of-domain), our model is on par with other dense retrieval models.
On BEIR, the dense retrieval models perform worse than BM25, which is consistent with existing work.
Overall, these experiments verify that our cosDPR-distil model is at least on par with existing reference dense retrieval models.

Comparing the Anserini HNSW results with brute-force exact nearest neighbor search, we see that HNSW indexes are able to retain a high level of effectiveness.
We explore additional HNSW settings below.

\paragraph{Evaluation of indexing and search}
Table~\ref{table:search-with-different-indexing-configs} presents the results of a comparison between the Anserini (Lucene) and Faiss implementations of the HNSW indexing algorithm.
Performance was measured by varying the \texttt{efC} graph construction parameter, the number of threads used to build the index, and whether or not we optimize the Lucene index through segment merging.
Here, we fix the search parameter \texttt{efSearch}=1000 for all the index variants, focusing on the tradeoffs induced by the indexing parameters.

We first focus on the Anserini (Lucene) implementation.
Overall, compared to Faiss FlatIP (brute-force nearest neighbor search), we see that the HNSW implementation successfully retains retrieval effectiveness while improving retrieval efficiency.
Additional observations:

\begin{itemize}[leftmargin=*]
\item Comparing rows (1) and (2) or (4) and (5), where the only difference is the number of threads used during indexing:\
we observe a significant reduction in indexing time, while achieving similar effectiveness in terms of MRR@10 and Recall@1k.
\item Comparing rows (1) and (4) or (2) and (5), where the only difference is the optimization of the index through segment merging:\ although optimization increases indexing time, segment merging brings significant improvements to query latency.
\item Comparing rows (2) and (3) or (5) and (6), where the \texttt{efC} parameter is increased from $100$ to $1000$:\ we see a slight improvement in effectiveness, but at the cost of significantly longer indexing times.
This comparison highlights the tradeoffs between indexing time and retrieval effectiveness.
\end{itemize}

\noindent When comparing the Anserini (Lucene) HNSW implementation with the Faiss implementation, the retrieval effectiveness achieved by both methods appears to be similar.
However, the indexing speed of Faiss is much faster than Anserini.

\begin{table*}[]
\centering
\begin{tabular}{ll|lllr}
\toprule
Row & Index    & efSearch & threads & MRR@10 & QPS \\
\midrule
(1)  & Anserini (HNSW) &   10        &    1          &   0.2551  & 587.8 \\
(2)  & Anserini (HNSW) &   100       &    1          &   0.3612  & 234.0 \\
(3)  & Anserini (HNSW) &   1000      &    1          &   0.3842  & 33.0  \\
(4)  & Anserini (HNSW) &   1000      &    16         &   0.3842  & 373.0 \\
\midrule
(5)  & Faiss (HNSW)    &   10        &    1          &   0.1906  & 1733.8 \\
(6)  & Faiss (HNSW)    &   100       &    1          &   0.3539  & 522.4  \\
(7)  & Faiss (HNSW)    &   1000      &    1          &   0.3806  & 60.5    \\
(8)  & Faiss (HNSW)    &   1000      &    16         &   0.3806  & 164.5  \\
\midrule
(9)  & Faiss (FlatIP)  &   -         &    1          &   0.3896  & 0.4     \\
(10) & Faiss (FlatIP)  &   -         &    16         &   0.3896  &  24.8  \\
\bottomrule
\end{tabular}
\vspace{0.25cm}
\caption{HNSW search with different search configurations; index was constructed with \texttt{M}=16, \texttt{efC}=100, single thread, optimized.}
\label{table:varying-search}
\vspace{-0.5cm}
\end{table*}

\paragraph{Further explorations of search}
In Table~\ref{table:varying-search}, we further explore how the search parameters influence the tradeoffs.
For index construction, we used \texttt{M}=16, \texttt{efC}=100, single thread, optimized. 
Here, we focus on varying two key parameters: the \texttt{efSearch} parameter and the number of threads used during the search.

For the Anserini (Lucene) implementation, comparing rows (1) to (3), we see that as the \texttt{efSearch} parameter increases, the MRR@10 metric also increases (slightly), reaching a peak of $0.3842$ at \texttt{efSearch}=1000, which is very close to exact nearest neighbor search.
However, QPS decreases significantly as the \texttt{efSearch} parameter increases, as expected.
Faiss exhibits the same behavior, shown in rows (5) to (7).

Compared to the HNSW implementation in Faiss, the Lucene implementation appears to be slower with a single query thread.
This can be seen in row (3) vs.\ row (7).
Effectiveness is comparable, but Faiss appears to be roughly twice as fast as Lucene.

For the Anserini (Lucene) implementation, we can see from rows (3) and (4) that increasing the number of threads from $1$ to $16$ while holding constant \texttt{efSearch}=1000 increases QPS from $33$ to $373$; multi-threaded search improves effectiveness, as expected.
The comparable Faiss results are shown in rows (7) and (8).
Interestingly, Faiss does not exhibit anywhere close to the increase in QPS from multi-threaded search that we observe in Lucene.

In summary, it appears that Lucene and Faiss effectiveness is comparable.
Using a single thread, Faiss is substantially faster, but Lucene appears to achieve better multi-threaded performance.
We are quick to concede that we have not sufficiently delved in the underlying implementation details to understand the source of these observed performance differences.

\section{Conclusion}

In this work, we explored the integration of Lucene's HNSW indexes in the open-source Anserini IR toolkit. Our approach provides a unified interface for both dense and sparse retrieval in Java, using only Lucene.
Experiments on the MS MARCO passage and BEIR datasets show that our cosDPR-distill model achieves reasonable effectiveness, but comparisons to the popular Faiss library reveal differences in performance that we have yet to fully explore.
Nevertheless, Anserini provides an intriguing value proposition:\ a ``single stack'', unified Lucene-based framework for {\it both} dense and sparse retrieval models, which potentially simplifies the deployment of production search applications that combine the best of both worlds in terms of the hybrid fusion of dense and sparse retrieval models.

\section*{Acknowledgements}

This research was supported in part by the Natural Sciences and Engineering Research Council (NSERC) of Canada.

\bibliographystyle{ACM-Reference-Format}
\bibliography{ref}

\end{document}